\documentclass[aps, prx, superscriptaddress]{revtex4-2}

\usepackage{graphicx}
\usepackage{amssymb,amsmath,amsfonts,latexsym, mathrsfs}
\usepackage{stmaryrd}
\usepackage{multirow}
\usepackage{xcolor}
\usepackage{soul}

\begin{document}

\title{FerroAI: A Deep Learning Model for Predicting Phase Diagrams of Ferroelectric Materials}

\author{Chenbo Zhang}
\email{cbzhang@tongji.edu.cn}
\affiliation{MOE Key Laboratory of Advanced Micro-Structured Materials, School of Physics Science and Engineering, Institute for Advanced Study, Tongji University, Shanghai 200092, China}
\author{Xian Chen}
\email{xianchen@ust.hk}
\affiliation{Department of Mechanical and Aerospace Engineering, Hong Kong University of Science and Technology, Clear Water Bay, Hong Kong}

\date{\today}

\begin{abstract}
Composition-temperature phase diagrams are crucial for designing ferroelectric materials, however predicting them accurately remains challenging due to limited phase transformation data and the constraints of conventional methods. Here, we utilize natural language processing (NLP) to text-mine 41,597 research articles, compiling a dataset of 2,838 phase transformations across 846 ferroelectric materials. Leveraging this dataset, we develop FerroAI, a deep learning model for phase diagram prediction. FerroAI successfully predicts phase boundaries and transformations among different crystal symmetries in Ce/Zr co-doped BaTiO$3$ (BT)-$x$Ba${0.7}$Ca$_{0.3}$TiO$_3$ (BCT). It also identifies a morphotropic phase boundary in Zr/Hf co-doped BT-$x$BCT at $x = 0.3$, guiding the discovery of a new ferroelectric material with an experimentally measured dielectric constant of 9535. These results establish FerroAI as a powerful tool for phase diagram construction, guiding the design of high-performance ferroelectric materials.
\end{abstract}

\maketitle

\section{Introduction}

Phase-transforming ferroelectric materials are extensively employed in functional devices, such as piezoelectric converters and pyroelectric generators \cite{liu2018role, zhang2021energy, bucsek2020energy}. Doping has long been considered an effective strategy for fine-tuning these phase transformations, as the transformation temperature and crystal structure are sensitive to the dopant's composition. Through systematic doping experiments with comprehensive property characterizations, the transformation temperature and crystal structure can be mapped, outlining the phase boundaries at varying compositions. This traditional route to construct a phase diagram is crucial because it highlights the conditions under which phase transformations occur. Recognition of the phase boundaries associated with chemical, thermal, and physical properties helps determine the effective dopant's composition to enhance material properties such as dielectric constant, piezoelectric coefficient, and pyroelectric coefficient \cite{xie2022structure, huang2024grain, dai2021phase}. The construction of precise and comprehensive phase diagrams not only reveals the fundamental effects of doping but also provides a rational strategy for the design of high-performance ferroelectric materials \cite{li2022high, wei2021ferroelectric, zheng2015strong, tian2023morphotropic}.

The typical phase diagrams of ferroelectrics, consisting of multiple metallic elements in one or two stoichiometric oxide compounds, are represented as a composition-temperature diagram between the two compounds. Phase boundaries are often determined by linear extrapolation of experimentally measured transition temperature of the compound with a varying composition of the selected element. Finer tuning of compositional parameters enhances the precision of the phase diagram, while a broader chemical tuning range improves its comprehensiveness. Theoretical approaches, such as phase-field simulations, are commonly employed for phase diagram construction, relying on semi-empirical thermodynamic parameters fitted from experimental data \cite{wang2020combining, sun2022thermodynamic, kroupa2013modelling}. However, predicting phase diagrams for complex materials, such as ferroelectric oxides with morphotropic phase boundaries (MPBs), remains challenging due to the lack of well-established thermodynamic parameters \cite{gorsse2018reliability}. Conventional methods based on experimental extrapolation and phase-field simulations become less reliable when applied to new material systems with unknown thermodynamic properties. With the increasing demand for novel ferroelectric materials in sensing, actuation, and energy applications, more accurate phase boundary prediction methods are essential for guiding material design. Therefore, alternative strategies that do not heavily depend on pre-existing thermodynamic data are needed to construct reliable phase diagrams for complex ferroelectrics.

Recently, artificial intelligence (AI) has emerged as a promising approach for constructing composition-temperature phase diagrams in ferroelectric materials. Pioneering studies have leveraged machine learning (ML) algorithms to predict crystal structures and phase transformation temperatures based on human-selected atomic features \cite{he2022machine, he2021machine, yuan2022machine}. While these models perform well for specific material systems, their predictive accuracy tends to decrease when applied to a broader range of ferroelectrics due to limited generalization. This limitation arises from the insufficient cross-material-family data available in current training datasets.

Although publicly available materials databases, such as the Materials Project \cite{jain2013commentary} and the Crystallography Open Database \cite{Vaitkus2023}, provide extensive information on chemical compositions, crystal structures, and electronic properties, phase transformation data is rarely available in these databases, specifically critical information such as temperature dependent transport properties, latent heat and symmetry relations among phases. Without a comprehensive and accurately labeled dataset of phase transformations, existing AI models are limited to specific material classes and face challenges in constructing wide-range phase diagrams that are broadly applicable to diverse ferroelectric systems. Thus, developing a general AI framework for phase diagram prediction requires not only robust models but also a well-curated dataset that captures the full complexity of phase transformations across various material families.

In this work, we construct a phase transformation dataset encompassing various ferroelectric materials and develop an AI model, \emph{FerroAI} to generate wide-range composition-temperature phase diagrams. The model is systematically optimized to enhance its generalization capability, demonstrating strong predictive performance across diverse ferroelectric systems. To validate its effectiveness, we compare the wide-range, high-resolution phase diagrams constructed by FerroAI, compared with discrete phase transformation data reported in the literature. Furthermore, we fabricate a new family of ferroelectrics and use experimental characterization to evaluate whether FerroAI can accurately predict their phase diagrams. The results confirm that FerroAI not only provides accurate predictions but also exhibits strong scalability in modeling phase transformations for unexplored materials, establishing it as a powerful and generalizable AI tool for ferroelectric materials discovery.

\section{Results}
\subsection*{Construction of phase transformation training set by text-mining}
A comprehensive dataset of ferroelectrics with consideration of phase transformations is essential for deriving the artificial intelligence models. However, comprehensive datasets covering a wide range of ferroelectric materials are scarce due to the limited connections among various symmetries of crystal structures for complex ferroelectric systems. To address this, we systematically compiled a large-scale dataset by extracting phase transformation information from published literature. By leveraging natural language processing (NLP) techniques, we identified key details such as chemical compositions, crystal structures, and transition temperatures from thousands of research articles. The detailed methodology of this text-mining process is described in Methods.

\begin{figure}[h]
    \centering
    \includegraphics[width=0.85\textwidth]{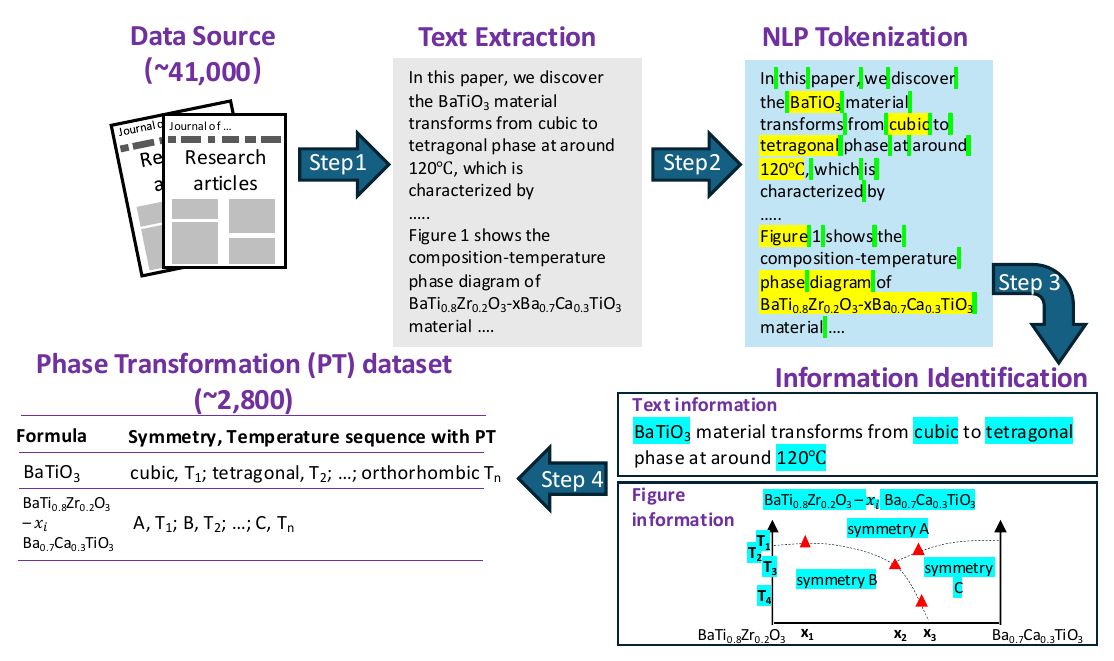}
    \caption{Data-mining process for ferroelectric materials with symmetry-breaking phase transformations by Nature Language Process (NLP). }
    \label{fig:mining}
\end{figure}

The dataset construction process is illustrated in Figure \ref{fig:mining}. The phase transformation dataset for ferroelectric materials was constructed by text-mining 41,597 research articles using Elsevier’s official Application Programming Interface (API). Through this process, we extracted phase transformation information, including chemical compositions represented by chemical formulas, crystal structures, and transformation temperatures associated with symmetry sequences across phase transitions.  

\begin{figure}[h]
	\centering
    \includegraphics[width=0.85\textwidth]{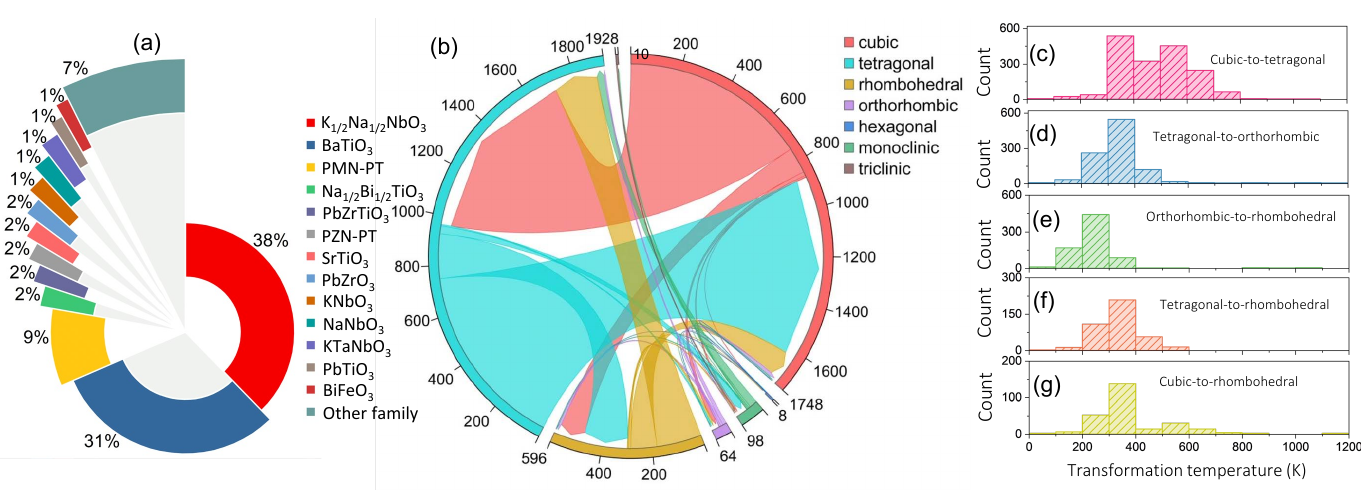}
	\caption{Statistics of classified ferroelectric phase transformation dataset: (a) data population from different families of doped ferroelectric materials; (b) chord diagrams of transformation relation between different crystal structures; (c)-(g) transformation temperature distribution in different types of phase transformation.} \label{Fig_statistics}
\end{figure}

After automated extraction and verification, we compiled a dataset comprising 2,838 phase transformations across approximately 800 ferroelectric materials. The distribution of ferroelectric materials in our dataset is illustrated in Fig.~\ref{Fig_statistics}(a).  Clustering analysis highlights that potassium sodium niobate, barium titanate, lead zirconate titanate, and lead magnesium niobate are the most extensively studied ferroelectric and piezoelectric materials in research publications on Elsevier. 

The relationships among 7 crystal systems for the materials in the dataset are visualized using a chord diagram, presented in Fig.~\ref{Fig_statistics}(b). It reveals that cubic to tetragonal and tetragonal to rhombohedral phase transformations are the most frequently observed transitions in ferroelectrics. The statistics of temperatures associated with the specific symmetry-breaking transformations are systematically shown in Figs.~\ref{Fig_statistics}(c)-(g). The histograms of these phase transformation temperatures indicates that most phase transitions occur within the range of 100~K to 700~K. Furthermore, the transformation temperature generally decreases as the symmetry of the crystal structure lowers.

\begin{figure}[h]
	\centering
	\includegraphics[width=0.65\textwidth]{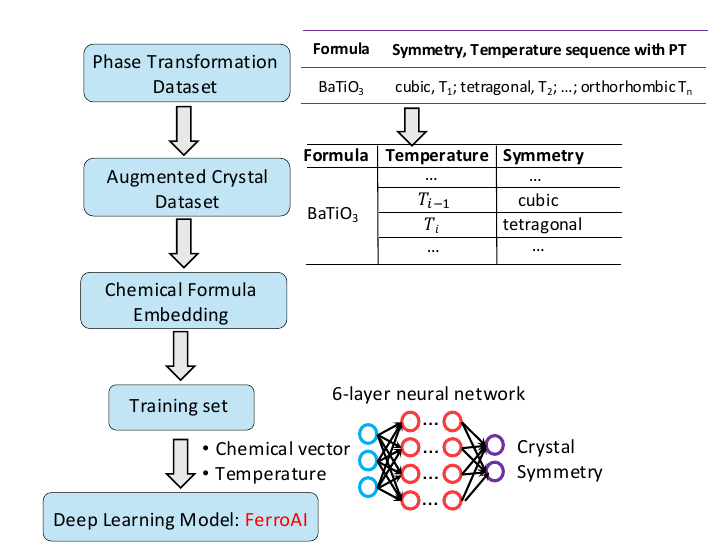}
	\caption{The workflow illustrating the construction of the augmented crystal dataset for deep neural network training and the development of the FerroAI model.} \label{Fig_method1}
\end{figure}

Since different material systems exhibit varying phase transformation sequences and multiple phase transitions commonly occur in a material among various symmetries, the phase transformation dataset (i.e. PT dataset) is further converted into a well-structured crystal dataset through data augmentation to enhance data consistency for training, as illustrated in Fig.\ref{Fig_method1}. 
To label the dataset, we assign symmetry labels to specific temperature ranges according to the phase transformation sequence in the PT dataset. We then introduce an augmentation factor, $N$, to uniformly divide each labeled temperature range into $N+1$ smaller intervals, extracting the boundary temperature points as input data. This ensures that all labeled temperature ranges are proportionally sampled, maintaining consistent relative weighting compared to the original phase transformation data.
Additionally, at every transformation temperature, we add data points corresponding to the crystal structures immediately before and after the transition, specifically at 1 K below the transformation temperature. This step improves the accuracy of phase boundary representation. Through these procedures, the original unstructured PT dataset is systematically converted into a well-structured, augmented crystal dataset.

\subsection*{FerroAI model for predicting phase diagram}

To develop the FerroAI model for phase diagram prediction, we used the deep learning neural network trained on the augmented crystal dataset. The overall process flow is illustrated in Fig.~\ref{Fig_method1}. In this framework, the input consists of the materials tagged by its chemical formula and atomic compositions. These tagged datasets serve as the foundation for training the neural network, allowing it to learn the compositional range of stable phases at specific temperatures. 

\begin{figure}[h]
	\centering
	\includegraphics[width=0.5\textwidth]{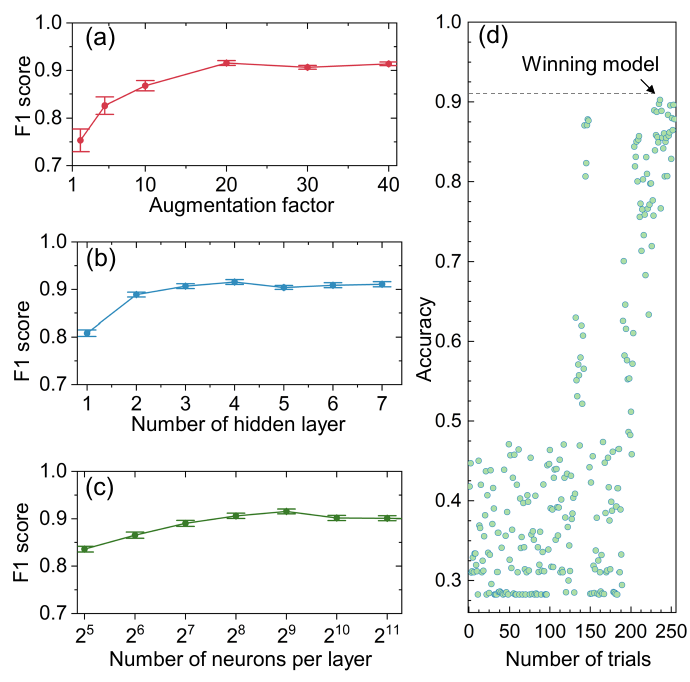}
	\caption{Optimization of primary hyperparameters: (a) augmentation factor; (b) number of hidden layer;(c) number of neurons per layer; (d) optimization of secondary hyperparameters via Hyperband approach.} \label{Fig_method2}
\end{figure}

We design a six-layer deep neural network with chemical vector and temperature as the input and the crystal symmetry as the output. The chemical vector refers to the material system through chemical formula embedding. Model performance is optimized by systematically tuning key hyperparameters, including the augmentation factor, number of hidden layers, neurons per layer and learning rate, using the controlled variable method. The predictive accuracy is evaluated using a weighted F1 score \cite{taha2015metrics}, which accounts for variations in dataset distribution across different crystal structures. Figs.~\ref{Fig_method2}(a)-(c) present the mean and standard deviation of the weighted F1 score from 10-fold cross-validation. The mean and standard deviation of the weighted F1 score for the tuning learning rate are included in Supplemental Materials. The optimal hyperparameter set, corresponding to the highest F1 score, is selected for model training. A summary of the model architecture and primary hyperparameters is provided in Table~\ref{tbl1}. The total number of parameters in the most optimized model is 811,015, which is sufficient to capture the influence of chemical compositions on phase transformation among wide range ferroelectric families.

\begin{table}[h]
\centering
\caption{Optimized architecture and list of primary hyperparameters in FerroAI.}\label{tbl1}
\begin{tabular}{cccc}
\hline
Layer &  Layer  & \multirow{2}{*}{Neurons} & \multirow{2}{*}{Activation}   \\ 
position & Type & &  \\\hline
1& Input  & As input & -\\
2& Dense  & 512      & ReLU \\
3& Dense  & 512      & ReLU \\
4& Dense  & 512      & ReLU \\
5& Dense  & 512      & ReLU \\
6& Output & As output & Softmax \\
\hline
\end{tabular}
\end{table}

The secondary hyperparameters, including the weight decay coefficient and dropout rate for each layer, are further optimized using the Successive Halving approach within the Hyperband algorithm \cite{jamieson2016non,li2018hyperband}. In this process, over 200 hyperparameter combinations are tested, and the one yielding the highest accuracy is selected for final model training. The model demonstrates robustness to variations in secondary hyperparameters, as multiple top-performing configurations yield comparable accuracy. More tuning results are given in the Supplemental Materials. The predictive capability of the model is primarily governed by the choice of primary hyperparameters listed in Table \ref{tbl1}. The activation functions used are ReLU \cite{ramachandran2017searching} for hidden layers and Softmax \cite{goodfellow2016deep} for the output layer.

\begin{figure}[h]
\centering
\includegraphics[width=0.4\textwidth]{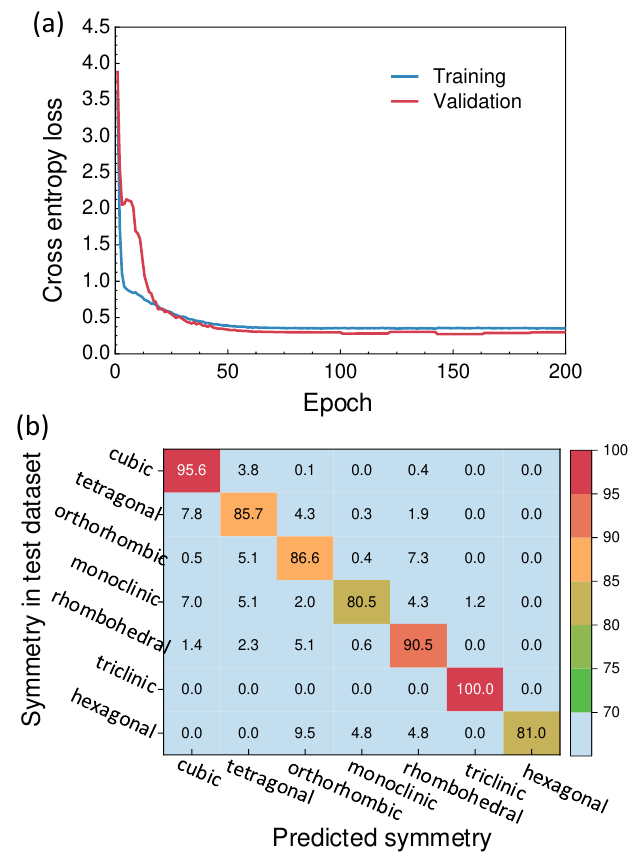}
\caption{Performance of FerroAI in training and testing: (a) cross entropy loss with training epochs corresponding to the best model; (b) confusion matrix for predicted crystal structure at unseen test dataset.} \label{Fig_model}
\end{figure}

\subsection*{Assessment of the FerroAI model for phase diagram prediction}

Using the optimal hyperparameters, we proceed with the final training of the model. Cross-entropy loss is used to evaluate model performance during training, quantifying the difference between the predicted and true probability distributions of the input data. The training objective is to minimize this loss. As shown in Figure \ref{Fig_model}(a), cross-entropy loss decreases rapidly with increasing training epochs, indicating that the model effectively learns from the input data. 

To assess the performance of our model in predicting phase diagrams for ferroelectric materials, we evaluate the trained model on a test dataset that was not used during training. The resulting confusion matrix, shown in Fig.~\ref{Fig_model}(b), represents the successful rate of crystal structure prediction. In this matrix, the horizontal axis denotes the crystal structures predicted by the FerroAI model at specific temperatures and compositions for the given ferroelectric materials, while the vertical axis represents the corresponding labeled structures from the test set. The FerroAI model demonstrates high accuracy in predicting crystal structures, particularly for cubic and rhombohedral phases. Although the model also achieves high accuracy for triclinic structures, the limited representation of triclinic data in the dataset constrains the reliability of this observation. 
A small fraction of tetragonal and orthorhombic structures are recognized as cubic or other symmetries, likely due to the marginal differences in their lattice parameters.
Overall, the model achieves over 80\% accuracy across all crystal structures, demonstrating that the well-trained FerroAI model effectively captures phase transformations and structural changes. The FerroAI model is available on Hugging Face, see Methods.

\begin{figure}[h]
	\centering
	\includegraphics[width=0.9\textwidth]{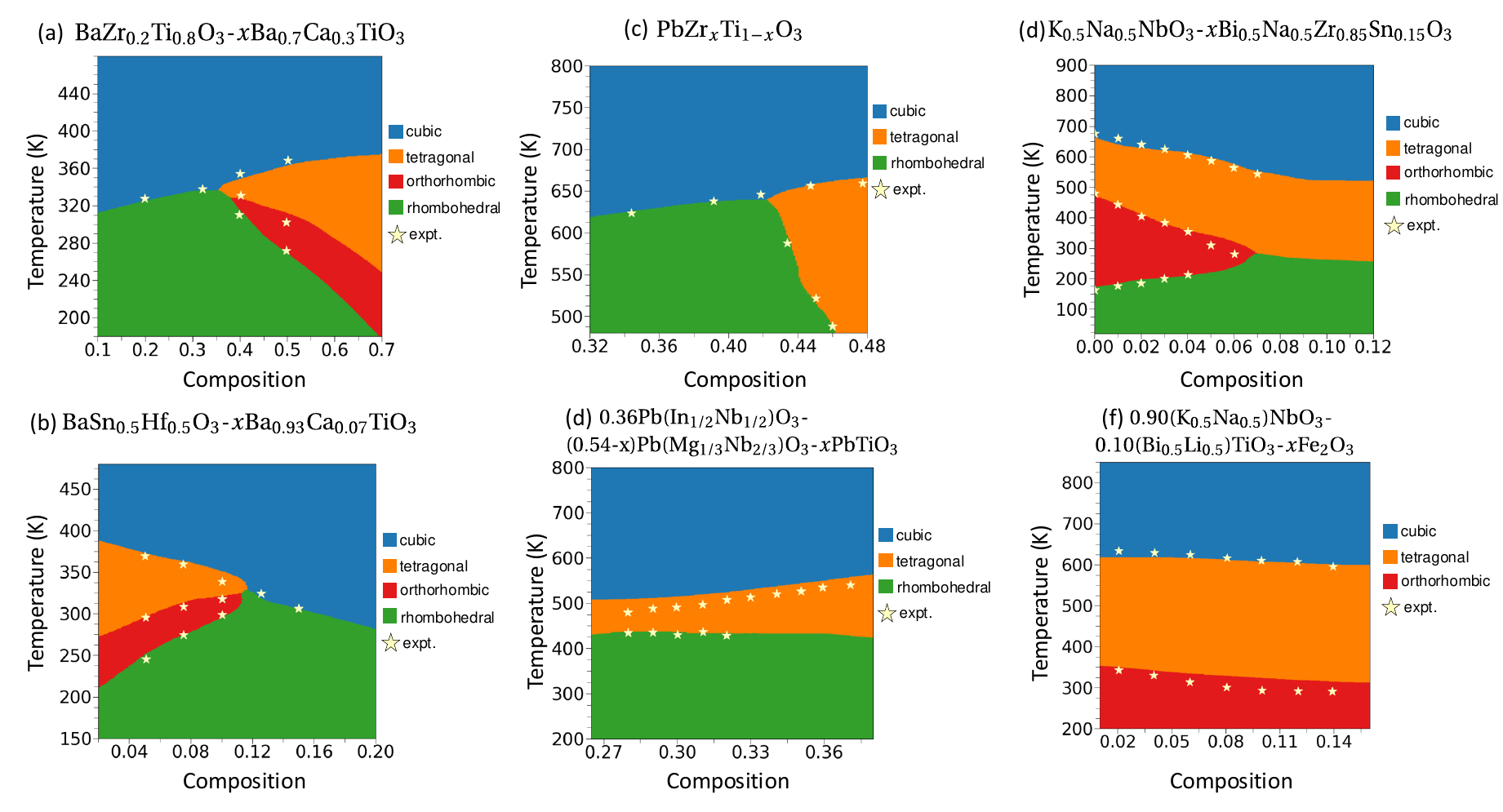}
	\caption{Phase diagrams predicted by FerroAI for ferroelectrics with respect to compositional variable $x$: (a) BaZr$_{0.2}$Ti$_{0.8}$O$_{3}$-$x$Ba$_{0.7}$Ca$_{0.3}$TiO$_{3}$ \cite{zhang2014phase}, (b) BaSn$_{0.5}$Hf$_{0.5}$O$_{3}$-$x$Ba$_{0.93}$Ca$_{0.07}$TiO$_{3}$ \cite{zhao2016phase}, (c)PbZr$_{x}$Ti$_{1-x}$O$_{3}$\cite{bouzid2005pzt}, (d) 0.36Pb(In$_{1/2}$Nb$_{1/2}$)O$_{3}$-(0.54-x)Pb(Mg$_{1/3}$Nb$_{2/3}$)O$_{3}$-$x$PbTiO$_{3}$ \cite{wang2012phase}, (e) K$_{0.5}$Na$_{0.5}$NbO$_{3}$-$x$Bi$_{0.5}$Na$_{0.5}$Zr$_{0.85}$Sn$_{0.15}$O$_{3}$ \cite{gou2018microstructure}, and (f) 0.90(K$_{0.5}$Na$_{0.5}$)NbO$_{3}$-0.10(Bi$_{0.5}$Li$_{0.5}$)TiO$_{3}$-$x$Fe$_{2}$O$_{3}$ \cite{mahdi2020high}. } \label{Fig_demo}
\end{figure}

To illustrate the effects of doping elements on phase transformations in ferroelectric materials, we utilize FerroAI to generate high-resolution phase diagrams for common ferroelectrics, as shown in Fig.~\ref{Fig_demo}. The phase diagrams are constructed with a compositional resolution of 0.01 at.\% and temperature resolution of 1~K, enabling detailed visualization of phase boundaries. 
In the predicted phase diagrams, different colors represent distinct crystal symmetries, delineating phase boundaries at specific temperatures and compositions. The phase boundaries and crystal structures predicted by FerroAI align closely with experimental data extracted from the literature. Here, the discrete experimental data points \cite{zhang2014phase, zhao2016phase, bouzid2005pzt, wang2012phase, gou2018microstructure, mahdi2020high} are in the test dataset. The consistency between the predicted and experimentally observed composition-temperature phase diagrams demonstrates that FerroAI effectively captures the impact of compositional variations on phase transformations across diverse ferroelectric families.

Figs.~\ref{Fig_demo}(a)-(b) presents phase diagrams for complex barium titanate-based compounds. For both BaZr$_{0.2}$Ti$_{0.8}$O$_{3}$-$x$Ba$_{0.7}$Ca$_{0.3}$TiO$_{3}$\cite{zhang2014phase} and BaSn$_{0.5}$Hf$_{0.5}$O$_{3}$-$x$Ba$_{0.93}$Ca$_{0.07}$TiO$_{3}$\cite{zhao2016phase},  the phase diagrams reveal morphotropic phase boundaries (MPB) among three polar phases: rhombohedral, orthorhombic, and tetragonal. Notably, the orthorhombic region is relatively narrower than the tetragonal and rhombohedral regions. The predicted phase boundaries align well with experimental data, demonstrating the accuracy of the FerroAI model.

Fig.~\ref{Fig_demo}(c) presents the common piezoelectric PZT system with varying Zr composition from 0.3 to 0.5 at.\%. The MPB between the tetragonal and rhombohedral phases is accurately captured, showing good agreement with the literature \cite{bouzid2005pzt}.  
The phase diagram of complex Pb-based compounds is shown in Fig.~\ref{Fig_demo}(d). In this ternary system, the phase boundary remains insensitive to the composition of PbTiO$_3$ and does not exhibit an MPB. A similar phenomenon is observed in ternary potassium sodium niobates with varying Fe$_2$O$_3$, as shown in Fig.~\ref{Fig_demo}(f). In contrast, the binary potassium sodium niobate compound (Fig.~\ref{Fig_demo}e) exhibits an MPB among tetragonal, orthorhombic, and rhombohedral phases with the addition of a small atomic fraction of Bi$_{0.5}$Na$_{0.5}$Zr$_{0.85}$Sn$_{0.15}$O$_{3}$.

\section{Discussion}

To demonstrate the potential of FerroAI for discovery of new ferroelectric materials, we use it to predict phase diagrams for compositions not included in the training set, nor in test set. The supplemental video demonstrates a GUI to use the FerroAI model for phase diagram prediction. Specifically, we explore two ferroelectric systems: (1) Ba(Ce$_{0.005}$Zr$_{0.005}$Ti$_{0.99}$)O$_{3}$–$x$Ba$_{0.7}$Ca$_{0.3}$TiO$_{3}$ (abbreviated as BCeZrT–$x$BCT) and (2) Ba(Zr$_{0.1}$Hf$_{0.1}$Ti$_{0.8}$)O$_{3}$–$x$Ba$_{0.7}$Ca$_{0.3}$TiO$_{3}$ (abbreviated as BZrHfT–$x$BCT). While few compositions of BCeZrT–$x$BCT has been investigated for pyroelectric energy conversion \cite{zhang2023enhanced}, its phase diagram remains largely unexplored.

\begin{figure}[h]
	\centering
\includegraphics[width=0.8\textwidth]{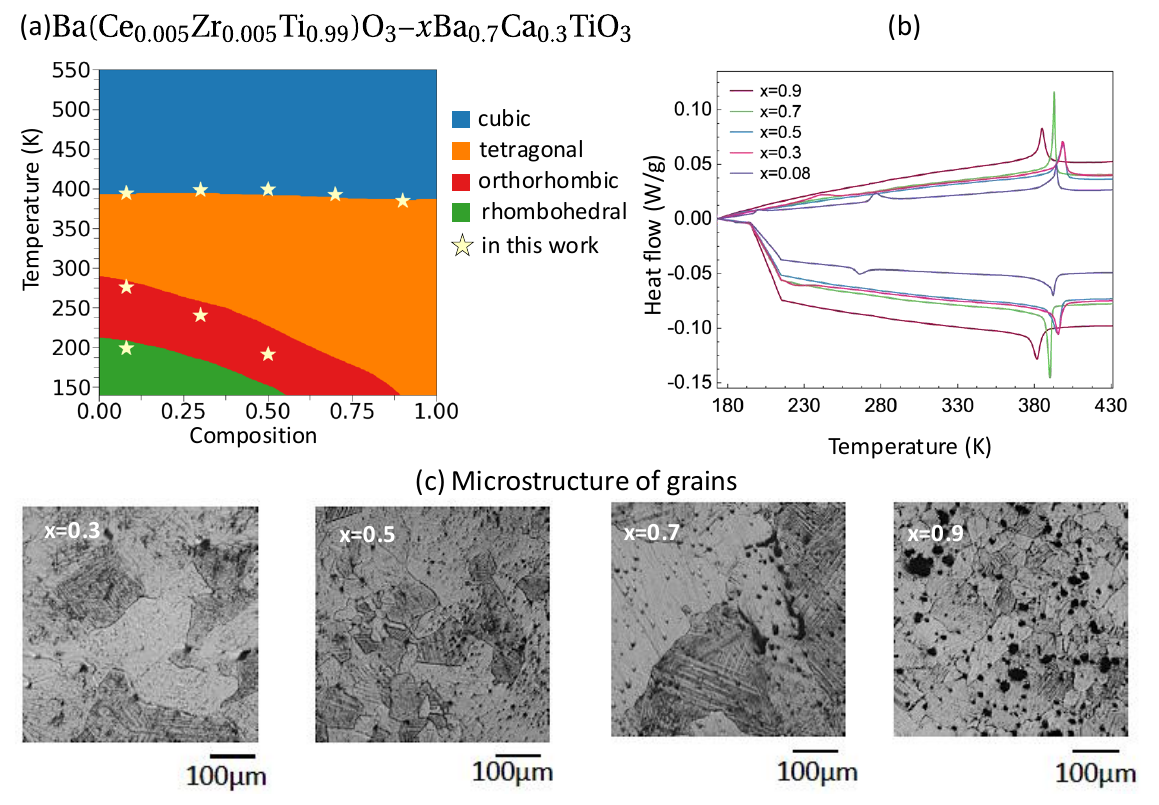}
	\caption{(a) Predicted phase diagram of synthesized ferroelectric material BCeZrT-$x$BCT, verified by (b) DSC measurement corresponding to the polycrystal grains observed in (c).} \label{Fig_expt}
\end{figure}

\subsection*{Phase transformations in binary compound Ba(Ce$_{0.005}$Zr$_{0.005}$Ti$_{0.99}$)O$_{3}$–$x$Ba$_{0.7}$Ca$_{0.3}$TiO$_{3}$}

The phase diagram for BCeZrT–$x$BCT is predicted across the full range of BCT compositions, as shown in Fig.~\ref{Fig_expt}(a), with a resolution of 0.01 at.\% and 1~K. The prediction reveals a sequential phase transformation from cubic to tetragonal, then to orthorhombic, and finally to rhombohedral between 550~K and 100~K. No MPB is observed in this binary compound system.  

To experimentally validate these predictions, we synthesized BCeZrT–$x$BCT samples with $x$ ranging from 0.08 to 0.9, following the synthesis method detailed in \cite{zhang2021energy}. Phase transformation temperatures were measured using differential scanning calorimetry (DSC) with a TA Instruments DSC 250, and the results are presented in Fig.~\ref{Fig_expt}(b). The extracted transformation temperatures are overlaid onto Fig.~\ref{Fig_expt}(a) for direct comparison. The FerroAI model accurately predicts the composition-dependent phase transformation sequence and the influence of BCT additions on phase transformation temperatures in this binary system. This result highlights FerroAI's capability in capturing complex thermodynamic behaviors in multicomponent ferroelectrics. Additionally, the microstructure of the synthesized samples, shown in Fig.~\ref{Fig_expt}(c), provides further insights into the room-temperature phases, revealing that samples with $x=0.3 - 0.7$ BCT exhibit a distinct grain morphology compared to those with $x=0.9$ BCT.

\subsection*{Prediction of morphotropic phase boundary for Ba(Zr, Hf)TiO$_3$-$x$Ba$_{0.7}$Ca$_{0.3}$TiO$_3$}

\begin{figure}[h]
    \centering
    \includegraphics[width=0.85\linewidth]{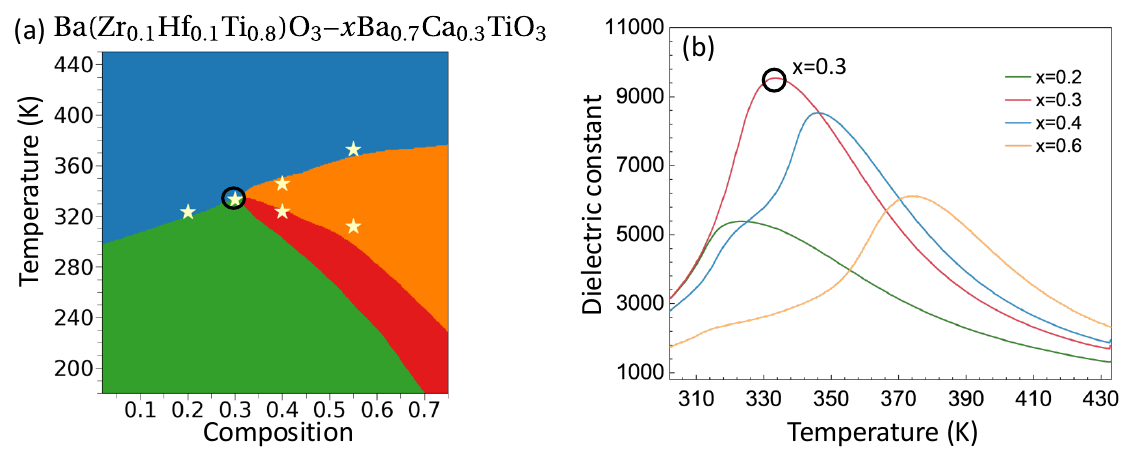}
    \caption{(a) Predicted phase diagram of BZrHfT-$x$BCT materials with $0.1\leq x\leq 0.7$, evaluated and verified by (b) the temperature dependent dielectric constant of selected compositions synthesized in this work.}
    \label{fig:ZrHf}
\end{figure}

Fig.~\ref{fig:ZrHf}(a) presents the FerroAI-predicted phase diagram for $0 \leq x \leq 0.7$ in BZrHfT–$x$BCT, using the same temperature and composition resolutions. The diagram reveals a sequential phase transformation from cubic to tetragonal, then to orthorhombic, and finally to rhombohedral (C–T–O–R). Different from the BCT doping in BCeZrT, this binary system exhibits MPBs among the three low-symmetry ferroelectric phases, emerging from $x=0.3$. This prediction has significant implications for the design of new ferroelectrics, as MPBs play a crucial role in enhancing functional properties for piezoelectric devices \cite{mishra2014enhanced, yao2012large, zhang2022enhanced}.

To verify the prediction, we synthesize BZrHfT-$x$BCT samples with $x$ varying from 0.2 to 0.55 near the predicted MPB. We measure the temperature-dependent dielectric constants using the aixACCT TF2000E ferroelectric analyzer at 1 kHz, as shown in Fig. \ref{fig:ZrHf}(b). The results indicate that at the predicted MPB, the $x=0.3$ sample exhibits a dielectric constant of 9535 at the transformation temperature, which is significantly higher than other compositions.  
We extract the transformation temperatures and plot them as markers in Figure \ref{fig:ZrHf}(a), showing strong agreement with FerroAI predictions. 

We summarize the MPBs predicted by FerroAI and confirmed through experimental characterization for various lead-free and lead-based materials in Table \ref{tbl2}. The comparison shows that FerroAI accurately predicts the MPB types, and the emerging compositions align well with experimental results. These findings demonstrate that FerroAI serves as a fast and reliable tool for identifying MPB compositions, providing valuable guidance for designing high-performance phase-transforming ferroelectric materials.  

\begin{table}
\centering
\caption{Comparison between FerroAI predicted morphological phase boundary and experimental result in lead-free and lead-based ferroelectric materials. }\label{tbl2}
{\small
\begin{tabular}{ccccc}
\hline
\multirow{2}{*}{Material} & \multirow{2}{*}{Type of MPB} & \multicolumn{2}{c}{Composition ($x$)}  & \multirow{2}{*}{Ref.}  \\
\cline{3-4}
& & FerroAI & Experiment & \\
\hline
BaZr$_{0.2}$Ti$_{0.8}$O$_{3}$-$x$Ba$_{0.7}$Ca$_{0.3}$TiO$_{3}$ & C-T-O-R & 0.35 & 0.35 & \cite{zhang2014phase}\\
BaSn$_{0.5}$Hf$_{0.5}$O$_{3}$-$x$Ba$_{0.93}$Ca$_{0.07}$TiO$_{3}$ & C-T-O-R & 0.11 & 0.1$\sim$0.12 & \cite{zhao2016phase}\\
PbZr$_{x}$Ti$_{1-x}$O$_{3}$& T-O-R &0.05&0.05&\cite{bouzid2005pzt}\\
K$_{0.5}$Na$_{0.5}$NbO$_{3}$-$x$Bi$_{0.5}$Na$_{0.5}$Zr$_{0.85}$Sn$_{0.15}$O$_{3}$& T-O-R&0.07&0.05&\cite{gou2018microstructure}\\
BaZr$_{0.1}$Hf$_{0.1}$Ti$_{0.8}$O$_{3}$-$x$Ba$_{0.7}$Ca$_{0.3}$TiO$_{3}$& C-T-O-R&0.30&0.30& this work\\
\hline
\end{tabular}
}
\end{table}

\section{Conclusion}
In conclusion, we have developed a deep learning neural network, FerroAI, trained on a comprehensive phase transformation dataset of ferroelectric materials. Leveraging 2,838 phase transformations from approximately 800 ferroelectric materials with reliable references, FerroAI effectively captures the influence of dopants on phase transformations and enables the rapid construction of wide-range composition-temperature phase diagrams.  

To validate its accuracy, we synthesized and characterized two ferroelectric material systems, demonstrating strong agreement between FerroAI-predicted phase diagrams and experimentally measured transformation temperatures. Based on the predicted MPB, we discovered that the BZrHfT-$0.3$BCT material exhibits a dielectric constant of 9535 at its phase transformation, significantly surpassing neighboring compositions.  These findings establish FerroAI as not only a robust tool for generating wide-range composition-temperature phase diagrams but also a valuable resource for guiding the design of high-performance phase-transforming ferroelectric materials.  

\noindent {\bf Acknowledgments} C. Z. acknowledge the support by National Natural Science Foundation of China (No.12204350), Fundamental Research Funds for the Central Universities (No.22120240095). X. C. thank the financial support under GRF Grants 16203021, 16204022, 16203023 by Research Grants Council, Hong Kong. 

\section*{Methods}\label{sec:method}
\subsection*{Training Dataset}
We extract the main texts from each of mined articles, removing references and other irrelevant sections. We use  Spacy library to process paragraphs and captions of figures and extract core phrases, which are enumerated for information identification. We define a list of rules to regularize the expressions to identify key information as seen in Step 3 of Fig.\ref{fig:mining}. Finally, the information including chemical formula, symmetry and associated temperature sequence is compiled to phase transformation dataset.

\subsection*{Material characterization}
The micrographs of synthesized samples were observed by the polarized light-reflected differential interference microscope\cite{zeng2019quantitative}. All samples were well polished by 1$\mu$m diamond suspension and etched by 37\% Hydraulic acid to reveal grain boundaries. The temperature dependent heat flow was measured under Nitrogen atmosphere. The heating and cooling rate is 7$^\circ$C/min. The temperature dependent dielectric constant is measured by the aix ACCT TF2000E ferroelectric analyzer via the Capacitance-Voltage module. The temperature step is 1$^\circ$C.

\subsection*{Availability of FerroAI model}
The best-trained model, FerroAI, along with the temperature and embedded elementary vector scaler, is available on Hugging Face at huggingface.co/FerroAI/FerroAI, facilitating broader accessibility and further exploration of phase diagram predictions for ferroelectric materials. The supplemental video demonstrates how to generate a phase diagram by providing an appropriate chemical vector input and temperature range.

\noindent {\bf Data declaration:} The data from figures to support this study are available on reasonable request from the corresponding author. 

\bibliographystyle{apsrev4-2}
\bibliography{cas-refs}

\end{document}